\documentclass[sigconf]{acmart}
\AtBeginDocument{%
  }

\usepackage{graphicx} 
\usepackage{multirow}
\usepackage{framed}
\usepackage{xcolor}
\usepackage{balance}

\copyrightyear{2026}
\acmYear{2026}
\setcopyright{cc}
\setcctype{by-nc-nd}
\acmConference[ICSE-FoSE]{2026 IEEE/ACM 48th International Conference on Software Engineering: Future of Software Engineering }{April 12--18, 2026}{Rio de Janeiro, Brazil}
\acmBooktitle{2026 IEEE/ACM 48th International Conference on Software Engineering: Future of Software Engineering (ICSE-FoSE), April 12--18, 2026, Rio de Janeiro, Brazil}
\acmPrice{}
\acmDOI{10.1145/3793657.3793895}
\acmISBN{979-8-4007-2479-4/2026/04}

\begin{document}

\title{Be a Partner, not a Bystander in Software Engineering Practice: Bridging the Gaps between Academia and Industry}

\author{Mohammad Masudur Rahman}
\email{masud.rahman@dal.ca}
\orcid{0000-0003-3821-5990}
\affiliation{%
  \institution{Dalhousie University}
  \city{Halifax}
  \state{NS}
  \country{Canada}
}

\author{Mehil B. Shah}
\email{shahmehil@dal.ca}
\orcid{0009-0000-3584-0470}
\affiliation{%
	\institution{Dalhousie University}
	\city{Halifax}
	\state{NS}
	\country{Canada}
}

\renewcommand{\shortauthors}{Rahman and Shah}

\begin{abstract}
    \looseness=-1
Software engineering conferences bring together thousands of academicians and software practitioners so that academic research and professional practices can influence each other. In essence, a symbiotic relationship exists between the research community and the software industry, which must be maintained, nurtured and re-examined periodically. Given the major AI breakthroughs (e.g., LLMs) and large-scale adoption of AI by the software industry, a re-examination of the relationship between academia and the SE industry is highly warranted. In this position paper, we argue that the software engineering community is deeply concerned about its research impact and relevance to industry practices. By conducting an empirical study using the survey responses from the SE community, we not only provide compelling evidence supporting our position but also propose new calls for action and reforms in SE, and thus envision a new future for the software engineering community.
\end{abstract}

%

\maketitle

\section{Introduction}
Software Engineering (SE) was born as a branch of knowledge in response to the \emph{software crisis} \cite{software-crisis}, thanks to the initiative of the North Atlantic Treaty Organization (NATO) \cite{nato-se}. To establish any branch of knowledge as an engineering discipline, a dedicated community exercising, curating or extending such knowledge is required \cite{se-professionalism}. Toward achieving this goal, International Conference on Software Engineering (ICSE) was first organized in the Fall of 1975 \cite{se-history}, giving rise to a 50-year-old community. Since then, this conference along with other specialized ones (e.g., FSE, ASE, ICSME) has brought together thousands of brilliant minds from all over the world and has established a thriving community while enjoying significant relevance and interests. The community has made numerous scientific contributions towards the advancement of not only software processes, tools, methods, and practices but also the broader society in general \cite{icse-proc}. However, like other academic or professional communities, reflection or readjustment is needed to sustain the growth of our community \cite{comm-sustainability}.

Any professional practice without a theoretical foundation is not sustainable. Similarly, a theory without any practical application has a limited impact \cite{theory-practice}. Each year, SE conferences bring together thousands of academicians, researchers,  educators, students, and practitioners to achieve at least two major goals: 
(a) to influence the industry practices with in-depth empirical insights and theories from academia, and (b) to inform academic research and innovations, the professional insights and the challenges experienced by software practitioners. In other words, there exists a symbiotic relationship between academia and software industry \cite{symbiotic}, which must be recognized, maintained and nurtured for a sustained growth of the SE community. However, several AI breakthroughs (e.g., AlexNet \cite{alexnet}, Transformers \cite{transformer}) have redefined the landscape of computing in the last decade. Their derived technologies (e.g., Generative AI, Agentic AI) have also been widely adopted by the software industry in recent years. Given these extraordinary circumstances and emerging political realities (e.g., reduced funding from governments \cite{funding-cuts}), a deep reflection within the software engineering community and a new call to action involving major reforms and realignment are highly warranted.

In this position paper, we conduct an empirical study to better understand the relevance of SE researches to industry practices, the gaps between academia and the SE industry, and the potential ways to bridge such gaps. 
Based on our study findings, we propose the following thesis statement:
\begin{framed}
	\noindent
		\textbf{Thesis Statement.} Software engineering community is deeply concerned about its research impact and relevance to industry practices. Without addressing its gaps with the industry through a major reform, it might lose its remaining relevance and fail to make any meaningful impact.
\end{framed}

The rest of the paper is structured as follows. Section \ref{sec:method} discusses our study methodology, Section \ref{sec:evidence} offers evidence to support our argument, Section \ref{sec:counter} discusses counter-argument, and finally Section \ref{sec:call2action} concludes with a call to the SE community for concrete actions.

\section{Study Methodology}\label{sec:method}
\textbf{Research questions.} We ask three important research questions targeting our thesis statement. They help us collect relevant experimental data and guide our analysis. They are as follows.
\begin{enumerate}
	\item \textbf{RQ1}: Is software engineering community content or concerned about its research impact and relevance to practices?
	\item \textbf{RQ2}: What are the potential gaps between software engineering researches and industry practices?
	\item \textbf{RQ3}: How to bridge these gaps between academia and software industry? What are the calls to action?
\end{enumerate}

\textbf{Data collection.} To design our experiment,  we collect the anonymous survey responses \cite{fose2026-survey} captured and distributed by the organizing committee of the \emph{Future of Software Engineering} workshop \cite{fose2026}. Since we are interested in the theme -- \emph{impact and relevance to practices}, we target three questions from the distributed Excel document involving \emph{what works} (i.e., Q7), \emph{what does not work} (i.e., Q8), and \emph{what needs to change} (i.e., Q9). We also capture response metadata (e.g., respondent's profession) to gain additional context about the responses. We use popular Python libraries (e.g., \texttt{openpyxl}, \texttt{pandas}) to extract the response data from the Excel document.

\begin{figure}[!t]
	\centering
	\captionsetup{size=small}
	\includegraphics[width=3in]{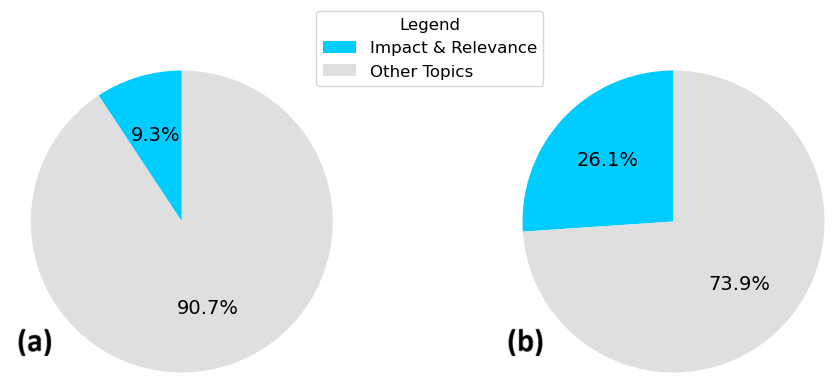}
	\vspace{-.2cm}
	\caption{Percentage of survey responses focusing on research impact and relevance to practices when discussing (a) what works, and (b) what does not work in SE}
	\label{fig:impact-discuss-ratio}
	\vspace{-1.6em}
\end{figure}

\textbf{Mixed method analysis.} The survey received 280 complete responses from the SE community. Since we are interested in the theme -- \emph{impact and relevance to practices}, we determine the pertinence of responses to the theme.
In particular, we manually check if they focus on any practical impact of SE researches or relevance to industry practices and annotate them as either \emph{yes} or \emph{no}. We employ two annotators and conduct our annotation in two rounds, with a discussion phase in between and at the end. We employ Cohen-Kappa agreement analysis and  achieve a \emph{near-perfect} agreement ($> 0.9$) between annotators in each of three response types. After resolving conflicts, we found that 26, 73, and 52 responses target our theme of interest when discussing what works, what does not work, and what needs to change in software engineering. 

We employ three frontier Large Language Models (LLMs) -- \emph{Claude (Sonnet 4.5)}, \emph{ChatGPT (GPT 5)} and \emph{Gemini 3 Pro} -- to the above response data to summarize their key concepts.
 Multiple models were chosen to mitigate the impacts of hallucination and the stochastic nature of LLMs' generated output \cite{multi-llm}. We carefully construct a prompt for each research question, instruct the model to focus only on our provided responses, and apply the same prompt to both AI models. Then we capture their verbatim responses for subsequent analysis and actionable insights.
The detailed experimental data, such as annotated responses, prompts, and LLMs' answers, are included in the replication package \cite{fose-replication} for replication and third-party reuse.


\begin{table*}
    \captionsetup{size=small}
	\caption{Responses expressing satisfaction and concerns about impact or relevance of software engineering research}
	\vspace{-.3cm}
	\label{table:content-concern}
	\resizebox{6.8in}{!}{
		\begin{tabular}{l|l||l|p{3.5in}}
			\hline
			\multicolumn{2}{c||}{\textbf{Content}} &  \multicolumn{2}{c}{\textbf{Concerns}} \\
			\hline
			\textbf{\#Responses} & \textbf{Theme} &  \textbf{\#Responses} & \textbf{Theme} \\
			\hline
			\hline
			\multirow{4}{*}{26 (9.07\%)} & Industry collaboration \& ties (34.62\%) & \multirow{4}{*}{73 (26.07\%)} & Content failures (28.76\%) : Research is not targeting the right problems.\\
			& Practical application \& tools (23.07\%) &  & Outcome failures (23.29\%): Research is not changing practices. \\
			& Real world validation \& relevance (15.38\%) &  & Structural failures (26.02\%) : Academia and industry operate separately.\\
			& Tangible impact on efficiency (7.69\%) &  & Incentive failures (34.24\%): Systems reward the wrong behaviours.  \\
			\hline
		\end{tabular}
	}
	\vspace{-.2cm}
\end{table*}

\section{Evidence for Thesis Statement}\label{sec:evidence}
In this section, we provide evidence to support our thesis statement by answering two of our research questions as follows.

\subsection{Answering RQ1: Expressed concerns} 
\begin{figure}[!t]
	\centering
	\includegraphics[width=3.5in]{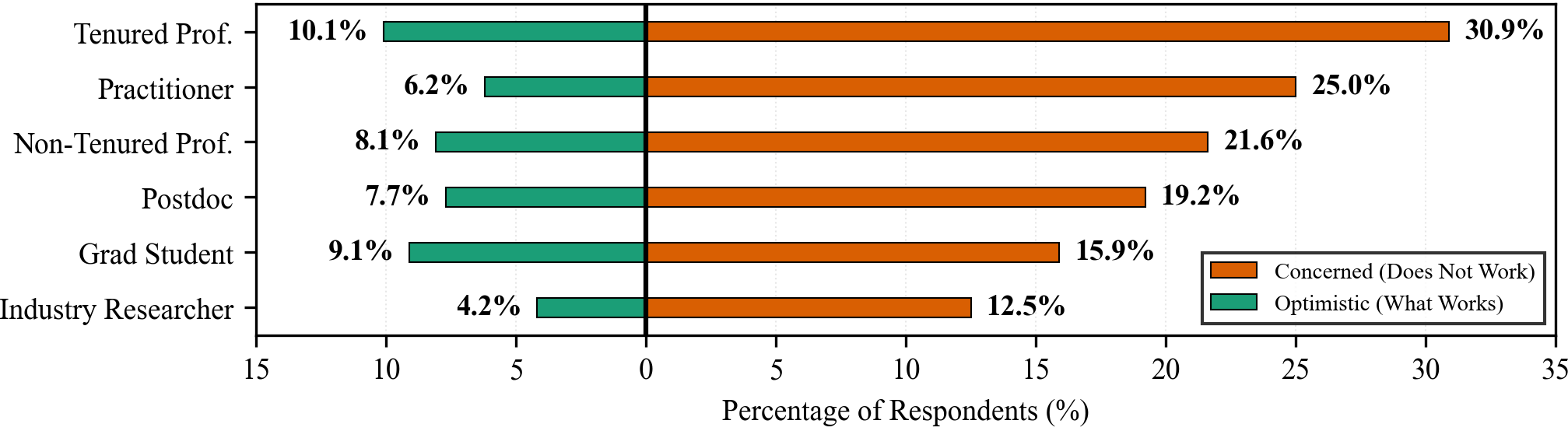}
	\vspace{-.6cm}	
	\caption{Percentage of respondents focusing on research impact and relevance to practices when discussing what works (left) and what does not work (right) in SE.} 
\label{fig:impact-ratio-demographics}
\vspace{-.5cm}
\end{figure}

Figures \ref{fig:impact-discuss-ratio} and \ref{fig:impact-ratio-demographics} summarize our investigation details for RQ1. To answer this research question, we analyze the survey responses to Q8, i.e., what does not work in SE. From Fig. \ref{fig:impact-discuss-ratio}, we see that 26.07\% (73) of 280 respondents are concerned about the impact or practical relevance of SE research, which is almost three times higher than the opposing views. We also analyze the demographics of respondents expressing such concerns. From Fig. \ref{fig:impact-ratio-demographics}, we see that tenured professors and practitioners are notably concerned about the impact and relevance. In fact, $\approx$31\% (46) of 149 tenured professors and 25\% (4) of 16 software practitioners discuss these issues in their responses. About $\approx$19\%  of non-tenured professors and postdoctoral fellows also voice similar concerns besides other issues.

We also manually analyze the survey responses about concerns (i.e., Q8) and LLMs' answers to our relevant prompt (a.k.a., Prompt 1a), and identify the key themes encompassing the concerns. It should be noted that the same survey response might target multiple themes. Table \ref{table:content-concern} summarizes our qualitative analysis. We found that the discussion focuses on four important themes as follows. 

\textbf{(a) Content failures.} According to $\approx$29\% of 73 responses, software engineering research does not target the right problems experienced by software practitioners. In other words, the research questions, dataset, and developed solutions might not match the real-world needs and contexts. Several respondents express their concerns as follows: ``Completely detached from practitioner reality'' (195868073), ``Too far from practice" (196091544), and ``Major disjoint between what is researched and what is useful" (195953547).

\textbf{(b) Outcome failures.} According to $\approx$24\% of 73 responses, SE research fails to make meaningful impacts on industry practices. In other words, academic research does not translate well into industry practices due to issues such as the use of toy datasets, the lack of industrial case studies, and poor maintenance of developed tools. Several respondents express their concerns as follows: ``Most industry practices do not come from software engineering research" (196290840), ``Industry is often ahead of the research community" (192305887),  ``Lack of impact on practice; few fundamental advances" (192676762), and ``...toy case studies rather than scale-realistic case studies  coming from genuine industrial needs" (196368894). Several responses also suggest the limited impact of SE research to policy, standards, and societal engagement.

\textbf{(c) Structural failures.} According to $\approx$26\% of 73 responses, SE research and practitioners communities exist in a parallel world with minimal interactions, communications, and mutual participation, leading to structural failures. Several respondents express their concerns as follows: ``Practitioners and researchers do not listen to each other" (192597855) and ``Very little meaningful industry presence at most SE conferences. It feels like we talk about industry all the time, but actual practitioners rarely attend" (196071446).

\textbf{(d) Incentive failures.} According to $\approx$34\% of 73 responses, academia rewards wrong behaviours, discouraging research of practical relevance. For example, \emph{publish or perish}, \emph{novelty over usefulness}, and \emph{quantity over quality} are commonly known issues. Several respondents express their concerns as follows: ``Work with practical relevance is undervalued, i.e., in separate tracks at conferences and journals that are valued less (treated as second class citizens)" (196065027),  ``Excessive focus on 'novelty' that encourages publishing very low-impact work...and aversion to work related to engineering/scalability/applicability that makes more real-world impact" (196380886), and ``... a broken peer-review system driven by incentives for high submission volume ..." (192101946).

\vspace{-2mm}
\subsection{Answering RQ2: Gaps between academia and SE industry} 
To better understand or explain the identified concerns above, we further analyze the survey responses (i.e., Q8) and LLMs' answers to our relevant prompt (a.k.a., Prompt 2), and investigate the gaps between SE research and software industry as follows.

\textbf{(a) Relevance and practical impact gap.}  Our analysis of the responses suggests that academic SE research is frequently disconnected from industrial needs and thus has minimal impact on practices. For example, one respondent suggests their concerns as follows: ``Most of the published SE research is irrelevant for practitioners...We improved x\% over a random and toy baseline is not something practitioners care about" (195853386).

\textbf{(b) Communication and engagement breakdown.} Researchers and practitioners operate in silos without meaningful interactions. Consequently, industry often advances independent of the research community. One representative response is as follows: ``Connecting research and practice. We both need to participate more into industry forums and to include them in ours academic ones. It is not about invitation, but actually a deeper synergy" (196181126).

\textbf{(c) Innovation direction gap.} According to the responses, the SE industry has been leading the innovations, whereas the research community follows trends. One plausible explanation is the lack of appropriate resources (e.g., computation, finance). For example, one representative response is: ``Most ideas that influence our field are not coming from SE research. SE research is mostly a follower, not an innovator (e.g., agile, micro-services, devops, etc.)" (196248189). 

\textbf{(d) Incentive misalignment.} Academia and industry have different reward structures. While academia focuses on publication metrics, the industry emphasizes software delivery and practical outcomes. For example, \emph{publish or perish} forces academicians to publish more and frequently rather than sustained, high-impact contributions. Similarly, emphasis on the novelty often encourages novel but low-impact and less applicable work to SE practices.

\textbf{(e)  Researcher experience gap.} Many academic researchers might lack industry work experience, limiting their ability to identify real industry needs. One representative response is as follows: ``Most SE researchers these days have zero industry experience ..., so it's hard for them to choose research that is both useful to industry and interesting to academia" (195910267).

\textbf{(f) Evaluation and validation gap.} Academic research might often use toy examples, outdated datasets, and unrealistic scenarios, limiting the applicability to industry-scale problems. One respondent states, ``Too many papers published evaluate the ideas they put forward on toy case studies rather than scale-realistic case studies coming from genuine industrial needs" (196368894).

\textbf{(g) Technology transfer gap.} There exist limited mechanisms and emphasis to translate SE research into industry practices. While there has been a major effort on the reproducible research (e.g., ROSE, ACM Badges), the knowledge transfer has not been smooth.

\textbf{(h) Focus and values gap.} Academia often values theoretical novelty, whereas the industry needs practical, maintainable, and scalable solutions, underscoring a fundamental gap between them.

\textbf{(i) Participation and representation gap.} The participation and representation of SE practitioners to academic conferences is limited, exacerbating the gap between the two communities. 

\looseness=-1
\textbf{(j) Resource and access gap.} Unlike the software industry, academia lacks computational and financial resources, limiting its capacity to work and innovate in certain cutting-edge technologies (e.g., LLMs). For example, one respondent expresses their concern as follows: ``Many recent SE studies rely on LLMs, but their high computational cost makes it difficult for researchers without industry collaboration to run large-scale experiments. As a result, companies with extensive GPU resources gain disproportionate influence over the direction and reproducibility of SE research" (195775809).

\textbf{(k) Knowledge dissemination gap.} Research findings remain confined within academic circles due to various issues (e.g., paywall) and hardly reach the practitioners. The outcomes of SE research are also not well communicated to the broader societies.

\textbf{(l) Research depth and continuity gap.} Academic research often targets short-term projects based on the availability of research grants and thus lacks depth and fails to develop lasting knowledge relevant to the industry. For example, one respondent suggests, ``Many research projects are short-term, driven by grants that last 2--3 years, which limits continuity'' (196070836).

Given our findings from \textbf{RQ1} and \textbf{RQ2}, we have strong evidence to believe that the SE community is genuinely concerned about its research impact and relevance to industry practices. 
Based on the survey responses, we also identify several inherent gaps between academia and the SE industry, contributing to such a challenge.

\section{Counter-arguments}\label{sec:counter}
Although the above analysis (Section \ref{sec:evidence}) provides strong evidence supporting our thesis statement, we also identify a set of responses that offer an optimistic view. In particular, we analyze the survey responses to Q7, i.e., what works in the SE community and the LLMs' answers to our relevant prompt (a.k.a., Prompt 1b) and capture evidence for formulating counter-arguments \cite{position-paper}. Figures \ref{fig:impact-discuss-ratio}, \ref{fig:impact-ratio-demographics}, and Table \ref{table:content-concern} summarize our analysis as follows.

From Fig. \ref{fig:impact-discuss-ratio}-(a), we see that $\approx$9\% (26) of 280 responses discuss positive impacts and relevance of SE research to industry practices. Fig. \ref{fig:impact-ratio-demographics}-(a) suggests that a maximum of 10\% of any respondent groups are optimistic about the impact of SE research. Our qualitative analysis in Table \ref{table:content-concern} also reveals several positive themes. For example, 9 out of 26 respondents express satisfaction in their collaboration with industry, industry-funded projects, and practitioner engagement. For example, one respondent suggests, ``Researchers frequently collaborate with companies like Google, Meta, Microsoft, Red Hat, and JetBrains. This will help them access large-scale real-world datasets, give the ability to validate research in practical settings, and help faster transfer of research innovations into tools and workflows." (195775809). Another response suggests a success story in empirical methods: ``The SE community has matured in its use of empirical and evidence-based methods...mining software repositories. Recent years have seen a push toward open science." (196070836). Another insightful response suggests the synergy between academia and industry as follows: ``The SE community works particularly well where it embraces the space between pure theory and pure implementation. This is where methodological pluralism, empirical evidence, and engineering practice can meet." (196113970).  At least 10 respondents were satisfied that their research can be easily integrated into SDLC, address real-world needs and thus makes tangible impacts (e.g., improving developer's efficiency), posing counter-evidence to our thesis statement.


\section{Conclusion \& Call to Actions}\label{sec:call2action}
Although the positive responses (Section \ref{sec:counter}) are encouraging, they are few in number and possibly submitted by the respondents involved with industry partners. 
On the contrary, evidence suggesting the limited impact of SE research and its weak relevance to industry practices is stronger, supporting our thesis statement. Given this finding, we analyze the responses to Q9, i.e., what needs to change and LLMs' answers to our relevant prompt (a.k.a., Prompt 3, RQ3), and formulate our seven-point call to action.

\textbf{Action 1: Publication and conference reform.} To change the status quo, the publication culture and conference organization should be significantly reformed. For example, reduction of publication pressure could lead to thoughtful, insightful research contribution. Reducing the number of submissions per author and increasing the number of reviewers per paper could also produce high-quality reviews, benefiting the papers. Merging technical and industry tracks in SE conferences and joint academic-industry conferences could help mitigate the widening gaps between academia and industry. Similarly, emphasis on novel ideas should be re-calibrated to encourage their impact on SE practices.

\textbf{Action 2: Mandatory industry engagement.} To gain a deeper insight into SE practices, researchers might need experience working in the industry. According to survey responses, SE researchers should work in the industry periodically (e.g., one year every decade). They should also use their research outcomes in the actual industrial software development, demonstrating their applicability.

\textbf{Action 3: Communication and outreach reform.} The SE community needs a major reform in their communication and outreach strategies. Public-facing mechanisms such as SE podcast or technology radar (e.g., Thought Works Technology Radar) can be employed to highlight exceptional, high-impact research outcomes to the general public. Short summaries of research papers and recorded talks can be shared on social media to engage more non-academics, which could breed new ideas. The SE community should also rethink its impact on societal issues and bridge the long-standing gaps with other research communities. 

\textbf{Action 4: Industry participation and partnerships.} Several major reforms are needed to encourage more industry participation and partnership in SE research. For example, according to the responses, software practitioners should be empowered with influential roles in SE conferences, rather than simply with industrial chair positions, to encourage serious dialogues (e.g., so what questions) between academia and industry. Small-to-medium sized organizations should be given enough opportunities to participate and contribute to the scientific discussions. Collaboration barriers should be mitigated through less proposal writing overhead and allowing more freedom to collaborate.

\textbf{Action 5: Research focus readjustment.} The SE community should prioritize real industrial needs and problem-solving. A balance between genuine inquiry for knowledge with industry's cost reduction needs must be established for the long-term sustainability of the community. SE should also strive to reposition itself as a strategic discipline to contribute to influential areas such as digital sovereignty, networking, sustainability, and trustworthy AI. Several relevant topics to the industry, such as requirements or architecture, deserve more attention from the SE community. Besides empirical SE, there should be more focus on developing high-value tools \& durable artifacts. Such efforts can be recognized through formal awards such as ``Most Impactful Tool Award".

\textbf{Action 6: Evaluation \& quality standards.} Traditional measures of success (e.g., h-index, citations) should be critically examined since they might be loosely connected to practical impact and actual problem solving. SE researchers might need to adopt successful industry practices, such as \emph{ship fast and then iterate} (e.g., the Y Combinator approach), to make their solutions effective.

\textbf{Action 7: Cultural and mindset changes.} According to survey responses, several common practices should be discouraged, such as \emph{prestige games} within the community and chasing \emph{low hanging fruits}. Instead, the pursuit of problems relevant to SE practices and ideas with high impact should be cherished and promoted.



\bibliographystyle{ACM-Reference-Format}
\bibliography{sigproc}

\balance

\appendix
\section{Used Prompts}
\begin{framed}
\noindent
	\textbf{Prompt 1a}: We are trying to investigate if the community is concerned about the impact or practical relevance of software engineering research. Analyze each row from the document and provide statistical evidence of concerns expressed by the respondents.
	\vspace{1em}
	
	\noindent
	\textbf{Prompt 1b}: We are trying to investigate if the community is content about the impact or practical relevance of software engineering research. Analyze each row from the document and provide statistical evidence of content expressed by the respondents.
\end{framed}


\begin{framed}
	\noindent
	\textbf{Prompt 2}: We are trying to investigate the gaps between academic research and industry practices. 
	Analyze each row from the document and identify the key gaps between these two worlds.
	The analysis should only focus on the provided data only.
\end{framed}

\begin{framed}
	\noindent
	\textbf{Prompt 3}: What should we do differently to enhance the impact or practical relevance of software engineering research? How can we bridge the gaps between academia and industry? Analyze each row from the document and identify the unique ways suggested by the respondents. The analysis should be restricted to the document only.
\end{framed}

\end{document}